\documentclass[12pt]{article}

\usepackage{amsmath}

\usepackage[dvipdfmx]{graphicx}

\usepackage{subfig}

\usepackage{array}

\usepackage{amsfonts}

\usepackage{epsf}

\usepackage{epsfig}

\usepackage{amssymb}

\usepackage{bbm}

\usepackage{delarray}

\usepackage{caption}

\usepackage{amstext}

\usepackage{graphics}

\usepackage{epstopdf}

\usepackage[titletoc,toc,title]{appendix}

\usepackage{scrtime}

\usepackage[multiple]{footmisc}

\usepackage[square,numbers]{natbib}

\usepackage{color}

\makeatletter

\catcode`ð=\active
 \defð{\u{g}}
 \catcode`Ð=\active
 \defÐ{\u{G}}
 \catcode`Ý=\active
 \defÝ{\. I}
 \catcode`ö=\active
 \defö{\"{o}}
 \catcode`Ö=\active
 \defÖ{\"O}
 \catcode`ü=\active
 \defü{\"{u}}
 \catcode`Ü=\active
 \defÜ{\"{U}}
 \catcode`Þ=\active
 \defÞ{\c{S}}
 \catcode`þ=\active
 \defþ{\c{s}}
 \catcode`ý=\active
 \defý{{\i}}
 \catcode`ç=\active
 \defç{\d{c}}
 \catcode`Ç=\active
 \defÇ{\d{C}}

\renewcommand*\@fnsymbol[1]{\the#1}

\makeatother

\numberwithin{equation}{section}
\begin{document}

\title{\textbf{Bound-State Solutions of Dirac Equation for Kratzer Potential with Pseudoscalar-Coulomb Term}}

\author{{\small Altu\u{g} Arda\footnote{arda@hacettepe.edu.tr}\,\,}\\
{\small \emph{Department of Physics Education, Hacettepe University}}, \\
{\small \emph{06800, Ankara, Turkey}}\\{\small Ramazan Sever\footnote{sever@metu.edu.tr}\,\,}\\
{\small \emph{Department of Physics, Middle East Technical University}}, \\
{\small \emph{06800, Ankara, Turkey}}\\ \\}
\date{}

\maketitle

\begin{abstract}
We present exact analytical solutions of the Dirac equation in $(1+1)$-dimensions for the generalized Kratzer potential by taking the pseudoscalar interaction term as an attractive Coulomb potential. We study the problem for a particular (spin) symmetry of the Dirac Hamiltonian. After a qualitative analyse, we study the results for some special cases such as Dirac-Coulomb problem in the existence of the pseudoscalar interaction, and the "pure" Coulomb problem by discussing some points about pseudospin and spin symmetries in one dimension. We also plot some figures representing the dependence of the energy on quantum number, and potential parameters.


\end{abstract}
\newpage

\section{Introduction}

The quantum systems are investigated by solving the Schrödinger equation for lower velocities than the speed of light, and by solving the Klein-Gordon or the Dirac equation if one takes into account the relativistic effects. In particular, the Dirac equation has to be solved for including the spin effects. This equation is a milestone in the formulation of the relativistic quantum mechanics, and used in a wide range of physical sciences from high energy physics to quantum information [1], cavity quantum electrodynamics [2, 3], and quantum fluctuations [4].

The form of Dirac equation and its bounded solutions including the pseudoscalar term has received a special attention in literature. One reason is that the Dirac equation having also pseudoscalar potential can be handled as a Sturm-Liouville problem [5, 6]. The main symmetries of the Dirac equation written in terms of vector, scalar and pseudoscalar potentials are parity, chirality and charge conjugation [7-9]. Under the discrete chiral transformation which is related with $\gamma^5$ and this matrix is placed as a factor in pseudoscalar term in the Dirac equation, the sign of mass, scalar and pseudoscalar potentials are changed [7, 8].

In Ref. [10], the bound state solutions of the Dirac equation with pseudoscalar potential have been studied for a linear potential. In Ref. [6], the solutions with corresponding eigenfunctions have been presented with some unexpected results for a screened Coulomb potential. The investigation of the Dirac equation with a pseudoscalar term have been made for the Cornell potential in Refs. [9, 11]. The connection between spin and pseudospin symmetries has been presented also with the Dirac equation having scalar, vector and pseudoscalar terms for the case of the Coulomb potential [7]. In Ref. [8], the transmission coefficient and resonant state energies of the Dirac equation having also pseudoscalar term for the square potential have been studied and found that the bound states can exist for a critical value of pseudoscalar potential. Arda and co-workers have investigated the bound-state solutions with corresponding eigenfunctions of the Dirac equation for a mixed vector-scalar-pseudoscalar Hulth\'{e}n potential within the position-dependent mass formalism [12] where the results for some specific $q$-values and for some $PT$-symmetric forms of the potential have been also presented.

We tend to obtain the results for the bound-state solutions of the Dirac equation in ($1+1$)-dimensions for the generalized Kratzer potential by taking the pseudoscalar term as an attractive Coulomb potential. One may obtain two uncoupled, Schrödinger-like equations from the Dirac equation in $(3+1)\,D$ in the existence of a pseudoscalar potential having a quantum number in the term proportional to $1/r^2$ which is related with the pseudospin (spin) symmetry [13, 14]. It means that studying the bounded solutions for the Kratzer potential could be interesting. This potential corresponds to the "effective potential" appearing in the above uncoupled equations obtained only for the Coulomb potential. This is the reason to add a Coulomb term as a pseudoscalar one. In the meantime, the Schrödinger equation for the Kratzer potential is equivalent to that of a radial Coulomb problem with an effective value of rotational angular momentum [15]. The bound-state solutions of the Dirac equation for the pseudoscalar-Coulomb potential needs a critical analysis because the spinor is not an eigenfunction of the parity operator [16]. The Hamiltonian for a spin-$\frac{1}{2}$ particle including the terms of inverse-squared and inversely linear fields with an anomalous gyromagnetic ratio represents a dyon-system, and the same Hamiltonian similar to that of the MIC-Zwanziger system describing a spin-$0$ particle in the same fields [17].

The spin and pseudospin symmetries are important subjects within the nuclear theory. The relativistic Dirac Hamiltonian has some hidden symmetries. Within pseudospin symmetry, the total single-particle angular momentum splits into pseudo orbital and pseudo spin angular momentum, i.e., $\vec{J}=\tilde{\vec{L}}+\tilde{\vec{S}}$, so single-particle states with quantum numbers $(n, \ell, j=\ell+1/2)$ and $(n-1, \ell+2, j=\ell+3/2)$ are relabelled as pseudospin doublets : $(\tilde{n}=n-1, \tilde{\ell}=\ell+1, \tilde{j}=\tilde{\ell}\pm 1/2)$. Within the relativistic mean field, the term, which is $\frac{1}{E+2M^*-V}\,\frac{\kappa}{r}\,\frac{d(2M^*-V)}{dr}$, in the uncoupled, Schrödinger-like equations for spherical case is taken in connection with the spin-orbital potential, and in the absence of the pseudo spin-orbital term, which is $\frac{1}{E-V}\,\frac{\kappa}{r}\,\frac{dV}{dr}$, occurs the pseudospin symmetry (here, $E=\epsilon-M, V=V_V+V_S, M^*=M+V_S$, and $\kappa=-\ell-1$ for $j=\ell+1/2$, $\kappa=\ell$ for $j=\ell-1/2$ with the vector ($V_V$) and scalar ($V_S$) potentials). The pseudospin approximation is observed if $dV/dr=0$, but this condition is not satisfied in the nuclei and the pseudospin symmetry is an approximation. This approximation will be good under the condition that $\left|\frac{1}{E-V}\,\frac{\kappa}{r}\,\frac{dV}{dr}\right|\ll\left|\frac{\kappa(1-\kappa)}{r^2}\right|$. So the pseudospin symmetry is related with the competition between the centrifugal barrier and the pseudo spin-orbital potential [18]. The spin and pseudospin symmetries could be studied in ($1+1$)-dimensional Dirac equation since the Dirac equation in $(3+1)\,D$ with a mixing of spherically symmetric scalar, vector and tensor interactions could be turned into the one in $(1+1)\,D$ with a mixing of scalar, vector and pseudoscalar interactions when the spin-half particle is forced to move in one direction while the tensor interaction becomes a pseudoscalar one [19]. There is a continuous effort about these subjects in different systems such as stable, deformed, exotic and spherical nuclei, and about extending to include different perspectives such as perturbative study or SUSY approach to them [14].

In the present paper, we obtain the results for the bounded solutions of the Dirac equation in ($1+1$)-dimension for the generalized Kratzer potential by adding a pseudoscalar-Coulomb term for a particular symmetry. We plot some figures to support our analytical results. We present the results for the "pure" Coulomb problem in the existence of the pseudoscalar interaction. We give the solutions for the case where vector part of the potential is zero. We also study the analytical results for the non-relativistic limit where the dependence of the energies on the pseudoscalar term changes slightly. We give our conclusions in last Section.

\section{Dirac Equation in $1+1$ Dimensions}

Time-independent Dirac equation describing a spin-$1/2$ particle, including scalar ($V_{S}(x)$), vector ($V_{V}(x)$) and pseudoscalar ($V_{P}(x)$) potentials can be written in terms of $\Sigma=V_{V}(x)+V_{S}(x)$, and $\Delta=V_{V}(x)-V_{S}(x)$ as ($\hbar=c=1$) [5-10, 20-24]
\begin{eqnarray}
\left\{\sigma_{1}p+\sigma_{3}[M+\frac{1}{2}\,(\Sigma-\Delta)]+\frac{1}{2}\left(\Sigma+\Delta\right)+\sigma_{2}V_{P}-E\right\}\Psi(x)=0\,,
\end{eqnarray}
where $\sigma_{1}, \sigma_{2}$ and $\sigma_{3}$ are the Pauli spin matrices.

The one-dimensional Dirac equation is covariant under $x  \rightarrow -x$ if pseudoscalar potential changes sign while vector and scalar part of potential remain unchanged. The sign of the energy, of the pseudoscalar and of the vector potentials change under the symmetry operation of charge-conjugation. So, the Hamiltonian is not invariant under this symmetry when it contains vector and pseudoscalar potentials. The other important operation for the Dirac Hamiltonian is the discrete chiral transformation under which the sign of the mass, of the scalar and of the pseudoscalar potentials are changed [7].

By taking the Dirac spinor as $\Psi=(\phi_{1}, \phi_{2})^{t}$ where $t$ indicates the transpose, we obtain the following first order coupled equations for upper and lower components
\begin{eqnarray}
&&\left(\frac{d}{dx}-V_{P}\right)\phi_{1}(x)=i[E+M-\Delta]\phi_{2}(x)\,,\\
&&\left(\frac{d}{dx}+V_{P}\right)\phi_{2}(x)=-i[-E+M+\Sigma]\phi_{1}(x)\,.
\end{eqnarray}

Writing $\phi_{2}(x)$ in terms of $\phi_{1}(x)$ with the help of Eq. (2.2), and inserting it into Eq. (2.3) gives us
\begin{eqnarray}
-\frac{d^{2}\phi_{1}(x)}{dx^{2}}+\left[(E+M)\Sigma+V^2_{P}+\frac{dV_{P}}{dx}\right]\phi_{1}(x)=(E^2-M^2)\phi_{1}(x)\,,
\end{eqnarray}
and by following similar steps, we obtain the second order equation for lower component as
\begin{eqnarray}
-\frac{d^{2}\phi_{2}(x)}{dx^{2}}+\left[(E+M)\Sigma+V^2_{P}-\frac{dV_{P}}{dx}\right]\phi_{2}(x)=(E^2-M^2)\phi_{2}(x)\,.
\end{eqnarray}
where we write last two equations for the case where $\Delta=0$ meaning that the Dirac equation has the spin symmetry [18, 19] for $E \neq M$. The coupled equations in  (2.2) and (2.3) have been well studied in [9] for $E=M$ where the authors obtained an extra solution for a particular potential not studied in [11].

In the next Section, we solve the above equations by identifying the sum of the scalar, and vector potentials as the generalized Kratzer potential and pseudoscalar potential by taking as an attractive Coulomb potential.

\section{Bound-State Solutions}

The potentials in Eq. (2.4) can be written as
\begin{eqnarray}
\Sigma=-2D\left(\frac{a}{|x|}-\frac{1}{2}\,\frac{qa^2}{x^2}\right)\,\,;V_{P}=-\frac{b}{|x|}\,,
\end{eqnarray}
where $D$ is the dissociation energy, $a$ is related with the equilibrium internuclear distance, $b$ is a positive parameter, and the parameter $q$ identifies generalization of the Kratzer potential such that for $q=1$ the potential gives 'standard' Kratzer potential while $q=0$ corresponds to Coulomb potential [25]. This form of the Kratzer potential gives the opportunity in which we are able to study the results for the Dirac-Coulomb problem in the existence of the pseudoscalar potential. This form of the "effective" potential tells us that there are two separated contributions coming from the Coulomb-like parts related with the parameters $a$ and $b$. The term proportional to $1/x^2$ which is more singular than $-1/x$ could be seen as a perturbative term to the Coulomb potential which has exact solutions for the Dirac equation.

We take the problem on the half line because of the covariance of the Dirac equation under the symmetry operation $x \rightarrow -x$ (should be imposed boundary conditions on $\Psi(x)$ at the origin and at infinity, if necessary) [9]. The components of the Dirac spinor on the whole line with well-defined parities can be obtained with the help of $\Psi(x)$ defined on the half line [9]. So, inserting a new variable $y=2\sqrt{M^2-E^2\,}|x|$, and writing the upper component as $\phi_{1}(x) \sim e^{-y/2}y^{p}f(y)$ because of the normalizable asymptotic forms of the solution, one obtains
\begin{eqnarray}
y\frac{d^2f(y)}{dy^2}+(2p-y)\,\frac{df(y)}{dy}-\left(p-Da\,\sqrt{\frac{M+E}{M-E}\,}\,\right)f(y)=0\,.
\end{eqnarray}

In order to solve this equation, we use an algebraic equation
\begin{eqnarray}
p(p-1)-\xi^+=0\,,
\end{eqnarray}
with $\xi^+=D(E+M)qa^2+b(b+1)$ and
\begin{eqnarray}
p=\frac{1}{2}\,\left[1 \pm \sqrt{1+4\xi^+\,}\right]\,,
\end{eqnarray}
which helps us to clarify the required conditions for binding fermions in the Dirac equation: $\xi^+$, and hence the radicand in (3.4) can not be less than zero because of the bounding condition $-M<E<M$. The value of $\xi^+=-1/4$ is not valid because the potential parameters $q$ and $b$ are positive. So, there is just one possible value for $p$ as $p=\frac{1}{2}+\sqrt{1/4+\xi^+\,}$\,.

Eq. (3.2) has a form of the Kummer's equation (which is also known as confluent hypergeometric equation) [21]
\begin{eqnarray}
z\frac{d^2h(z)}{dz^2}+(c_2-z)\,\frac{dh(z)}{dz}-c_{1}h(z)=0\,.
\end{eqnarray}
Its solution is
\begin{eqnarray}
h(z)=AM(c_1, c_2, z)+Bz^{1-c_2}U(c_1, c_2, z)\,.
\end{eqnarray}
Here $A$ and $B$ are arbitrary constants and $M(c_1, c_2, z)$ can also denoted as $\,_{1}F_{1}(c_1, c_2, z)$ [26]. $M(c_1, c_2, z)$ and $U(c_1, c_2, z)$ are the confluent hypergeometric functions of first and second kind, respectively. We must set the arbitrary constant $B$ to zero in order to satisfy boundary condition on the wavefunction for $z \rightarrow 0$. The behaviour of confluent series for $|z| \rightarrow \infty$ is
\begin{eqnarray}
\,_{1}F_{1}(c_1, c_2, z) \rightarrow \frac{\Gamma(c_2)}{\Gamma(c_1)}\,e^{z}z^{c_1-c_2}\,,
\end{eqnarray}
which gives exponentially divergent wavefunction meaning that it could not be square integrable. In order to avoid this divergence, the confluent hypergeometric function must be cut off as confluent hypergeometric series which has finite terms meaning that $c_1=-n$ [26]. So, the solutions of (3.2) are given as
\begin{eqnarray}
f(y)=A \,_{1}F_{1}(p-Da\,\sqrt{\frac{M+E}{M-E}\,}, 2p, y)\,.
\end{eqnarray}
The above requirement implies that the solutions in Eq. (3.8) can be written in terms of the associated Laguerre polynomials $L_{n}^{2p-1}(y)$ [26], and also an energy eigenvalue equation
\begin{eqnarray}
\frac{M+E_n}{M-E_n}=\frac{1}{D^2a^2}\left[n+\frac{1}{2}+\sqrt{\frac{1}{4}+\xi^+\,}\right]^2\,,
\end{eqnarray}
with eigenfunctions including a normalization constant $N$
\begin{eqnarray}
\phi_{1}(y)=Ne^{-y/2}y^{\frac{1}{2}[1+\sqrt{1+4\xi^+\,}\,]}L_{n}^{\sqrt{(2b+1)^2+4D(E_n+M)qa^2\,}}(y)\,.
\end{eqnarray}

We follow the same steps to get the lower component of the Dirac spinor. For this case, we have the following algebraic equation for $p$
\begin{eqnarray}
p(p-1)-\xi^-=0\,,
\end{eqnarray}
 with $\xi^-=D(E+M)qa^2+b(b-1)$ and
\begin{eqnarray}
p=\frac{1}{2}\,\left[1 \pm \sqrt{1+4\xi^-\,}\right]\,.
\end{eqnarray}
Minimum value of $\xi^-$ is obtained for $E \rightarrow -M$ which gives us two values for $b$ as $b<1/2$ or $b>1/2$ in order to get a real value of the squared root. We have two regions for the parameter $b$ as $0<b<1/2$ or $b>1/2$ because $b$ should be positive. Under these restrictions, the possible value of $p$ is $p=\frac{1}{2}+\sqrt{1/4+\xi^-\,}$\,. The polynomial condition gives the following eigenvalue equation
\begin{eqnarray}
\frac{M+E_n}{M-E_n}=\frac{1}{D^2a^2}\left[n+\frac{1}{2}+\sqrt{\frac{1}{4}+\xi^-\,}\right]^2\,,
\end{eqnarray}
with eigenfunctions including a normalization constant $N$
\begin{eqnarray}
\phi_{2}(y)=Ne^{-y/2}y^{\frac{1}{2}[1+\sqrt{1+4\xi^-\,}\,]}L_{n}^{\sqrt{(2b-1)^2+4D(E_n+M)qa^2\,}}(y)\,.
\end{eqnarray}

The eigenenergies given in (3.9) and (3.13) are finite when $n \rightarrow +\infty$, and it is clear that the solutions for upper and lower components are symmetric under $b \rightarrow b+1$. The parameter $b$ in these equations, which controls the contribution of the pseudoscalar interaction to the energy, is placed under square root, so the contribution to the energy coming from the Coulomb-part of the potential is larger than the ones coming from the pseudoscalar interaction. Figs. (1) and (2) show the variation of energies given in Eqs. (3.9) and (3.13) for first three levels depending one some parameters. Fig. (3) represents the behavior of the upper and lower components of the Dirac spinor, $|\phi_1|^2$ and $|\phi_2|^2$, and the probability density, $|\phi_1|^2+|\phi_2|^2$. The plot shows that the quantum number $n$ determines the number of nodes of $\phi_1$ and $\phi_2$. Note from the figure that, for the ground state, the probability density increases towards to the "center" (means that $x \rightarrow 0$) while it decreases for higher $x$-values, whereas this situation is exactly opposite for two excited states. It is observed that $|\phi_1|$ is smaller than $|\phi_2|$ for ground state while $|\phi_2|$ is comparable to $|\phi_1|$ for excited levels. de Castro has discussed these results with the above restrictions for an "effective" Kratzer potential in details in the absence of the pseudoscalar term [27]. The author has analyzed also some further restrictions and achieved a common eigenvalue equation. In Figs. (1) and (2), wee see some intersection points which are nodes appearing in this one-dimensional system in an effective potential having the form of the Kratzer potential. The effect of the Coulomb part of the potential becomes more dominant while the value of $b$ increases in the region where the ground-state energy is grater than the others.

Now we tend to analyze the case of the Dirac-Coulomb problem in the existence of the pseudoscalar interaction term. For this aim, we set the parameter $q$ to zero in Eq. (3.1). For the upper component, we have $p=1+b$, and consequently the eigenvalue equation as
\begin{eqnarray}
E_n=M\,\frac{(n+1+b)^2-D^2a^2}{(n+1+b)^2+D^2a^2}\,,
\end{eqnarray}
with eigenfunctions
\begin{eqnarray}
\phi_{1}(y)=Ne^{-y/2}y^{1+b}L_{n}^{2b+1}(y)\,.
\end{eqnarray}

For the lower component, we have $p=b$, and the eigenvalue equation as
\begin{eqnarray}
E_n=M\,\frac{(n+b)^2-D^2a^2}{(n+b)^2+D^2a^2}\,,
\end{eqnarray}
with eigenfunctions
\begin{eqnarray}
\phi_{2}(y)=Ne^{-y/2}y^{b}L_{n}^{2b-1}(y)\,.
\end{eqnarray}
where $N$ is normalization constant in Eqs. (3.16) and (3.18).

It is seen that we can obtain the results for the lower component by setting the parameter $b$ as $b-1$ in (3.15) and (3.16). In order to get a non-negative value for the upper indices of the associated Laguerre polynomials in Eq. (3.18) the region for $b$-values $b>1/2$ has to be taken into account. It is observed that the Coulomb-part of the Kratzer potential has an increasing contribution to the energy while the one coming from the pseudoscalar interaction decreases with increasing $b$.

We are ready to present briefly the results for the case where we take the vector potential as $V_{V}(x)=0$. For this situation we have to set $\Sigma=V_{S}$, and $\Delta=-V_{s}$ in Eqs. (2.4) and (2.5). In order to get the upper component of the Dirac spinor one solves the following equation
\begin{eqnarray}
\left(\frac{d^2}{dx^2}-\frac{dV_{P}}{dx}-V^2_{P}-2MV_{S}+E^2-M^2\right)\phi_{1}(x)=0\,,
\end{eqnarray}
and the lower component can be obtained from the equation
\begin{eqnarray}
\phi_{2}(x)=\frac{1}{i(E+M+V_{s})}\left(\frac{d}{dx}-V_{P}\right)\phi_{1}(x)\,.
\end{eqnarray}

We restrict ourselves for a special value of the parameter $q$ as $q=0$ to present here our results. Using the transformation $y=2\sqrt{M^2-E^2\,}|x|$, and writing the new wavefunction as $\phi_{1}(x) \sim e^{-y/2}y^{p}f(y)$, one obtains
\begin{eqnarray}
y\frac{d^2f(y)}{dy^2}+(2p-y)\,\frac{df(y)}{dy}-\left(p-\frac{2DMa}{\sqrt{M^2-E^2\,}}\,\right)f(y)=0\,,
\end{eqnarray}
where we set $p(p-1)-b(b+1)=0$. Eq. (3.21) has a form of the Kummer's differential equation as in the case of Eq. (3.2). Due to similar analysis on the boundary conditions should be satisfied by the wave function we write the solution as
\begin{eqnarray}
f(y)=A\,_{1}F_{1}(p-\frac{2DMa}{\sqrt{M^2-E^2\,}}, 2p, y)\,.
\end{eqnarray}
This solution is finite if we write $p-\frac{2DMa}{\sqrt{M^2-E^2\,}}=-n$ which gives the energy spectrum as
\begin{eqnarray}
E_n=\pm\,M\sqrt{1-\frac{4D^2a^2}{(n+1+b)^2}\,}\,,
\end{eqnarray}
and the corresponding wave functions with a normalization constant $N$
\begin{eqnarray}
\phi_{1}(y)=Ne^{-y/2}y^{1+b}L_{n}^{2b+1}(y)\,,
\end{eqnarray}
where we have to chose $p=1+b$ from the algebraic equation because of the asymptotic behaviour of the wave function. The energy eigenvalues are finite when $n \rightarrow +\infty$, and these results are in consistent with the ones given by de Castro who analyzes the restrictions of the problem beyond the formal results [27].

We solve the following Schrödinger-like equation for investigating the contribution coming from the pseudoscalar interaction for the non-relativistic case [6]
\begin{eqnarray}
-\frac{d^{2}\phi_{1}(x)}{dx^{2}}+\left[2M(V_V+V_S)+V^2_{P}+\frac{dV_{P}}{dx}\right]\phi_{1}(x)=2M(E-M)\phi_{1}(x)\,,
\end{eqnarray}
in which vector and scalar parts of the potential are coupled to the mass while the pseudoscalar term does not a connection with them as expected. So, we could predict that the dependence of the bound-state energies on the parameter $b$ is different from the ones obtained for the relativistic case. Here, the other component is given as $\phi(x)_{2}=(p/2M)\phi_{1}(x)$.

Defining a new variable $y=2\sqrt{2M(M-E)\,}|x|$ in (3.25), writing the upper component of the Dirac spinor as $\phi_{1}(x) \sim e^{-y/2}y^{p}f(y)$ gives
\begin{eqnarray}
y\frac{d^2f(y)}{dy^2}+(2p-y)\,\frac{df(y)}{dy}-\left(p-\,\frac{\sqrt{2M\,}Da}{\sqrt{M-E\,}}\,\right)f(y)=0\,,
\end{eqnarray}
which has the eigenvalues
\begin{eqnarray}
E_n=M\left[1-2\left(\frac{2Da}{2n+1+\sqrt{(2b+1)^2+8MDqa^2\,}}\right)^2\right]\,,
\end{eqnarray}
which are finite when $n$ goes to infinity. The corresponding eigenfunctions are given
\begin{eqnarray}
\phi_{2}(y)=Ne^{-y/2}y^{\frac{1}{2}[1+\sqrt{(1+2b)^2+8MDqa^2\,}\,]}L_{n}^{\sqrt{(1+2b)^2+8MDqa^2\,}}(y)\,.
\end{eqnarray}
where $N$ is a normalization constant.

The eigenfunctions are obtained again in terms of the Laguerre polynomials while the bound state energies have no longer a linearly dependence on the parameter $b$, this is so because the vector part of the potential is not coupled to the energy of the whole system in the non-relativistic case. It is worth to say that the result given in Eq. (3.27) is similar to the one obtained in Ref. [17], since our Hamiltonian and the Hamiltonians used in Ref. [17] have the same structures. So, the problem in $3D$ stated in Ref. [17] could be seen as the one of a spin-$\frac{1}{2}$ particle moving in an effective one-dimensional Kratzer-type potential because of the pseudoscalar term in Dirac equation. In this reference, the authors explore an algebraic pattern constructed on these Hamiltonians, and call it a 'new, subtle' SUSY. The studying of the dynamical quantities within this 'new' supersymmetry could be generalized to the case where the Hamiltonian has an additional pseudoscalar term.

\section{Conclusions}

We have studied the analytical solutions of the Dirac equation in $(1+1)$-dimensions by taking the vector and scalar part of the potential as a generalized Kratzer potential and by taking the pseudoscalar term as an attractive Coulomb potential. The effect of a Coulomb-like potential taking as a pseudoscalar interaction on the Dirac eigenenergies are remarkable. We have presented the results for the Dirac-Coulomb problem, and also studied the results for the non-relativistic case. We have also visualized variation of energy values in a few plots which make our analytical results more clear.

\section{Acknowledgements}
We would like to thank the kind referee for positive suggestions which have improved deeply the present paper.

\newpage

\begin{figure}
\centering \subfloat[][$b=0.1$, $q=0.01$, $D=5$]{\includegraphics[height=2.2in,
width=3in, angle=0]{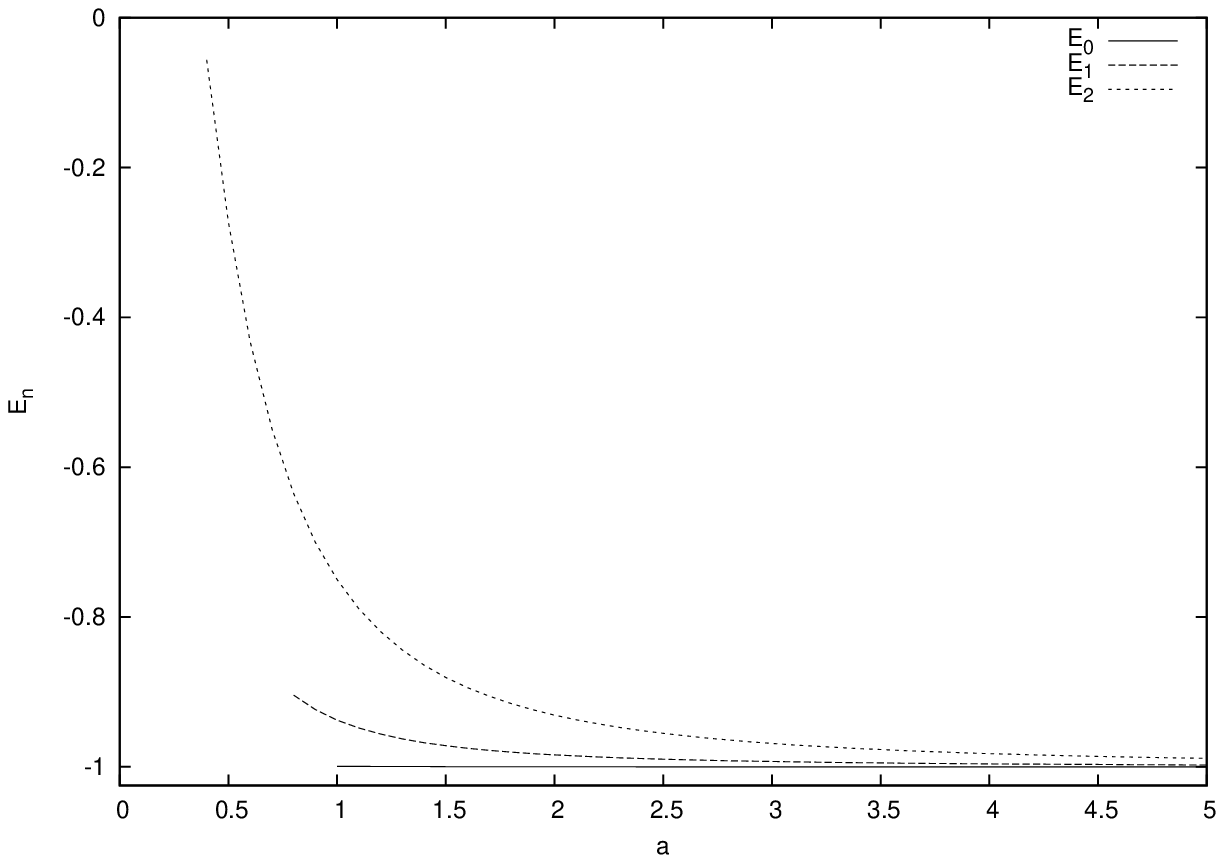}}
\subfloat[][$a=5$, $q=0.01$, $D=5$]{\includegraphics[height=2.2in, width=3in,
angle=0]{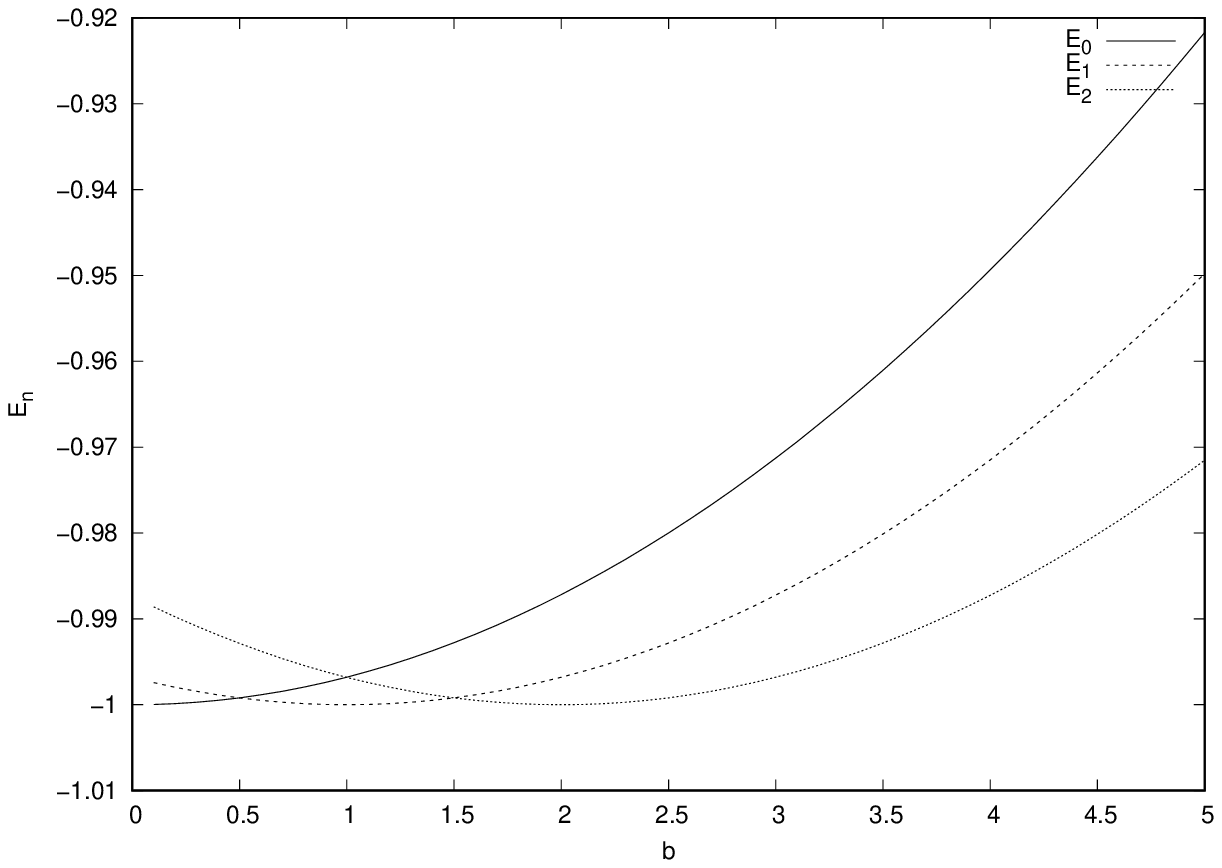}}\\
\subfloat[][$a=1$, $b=1$, $D=10$]{\includegraphics[height=2.2in,
width=3in, angle=0]{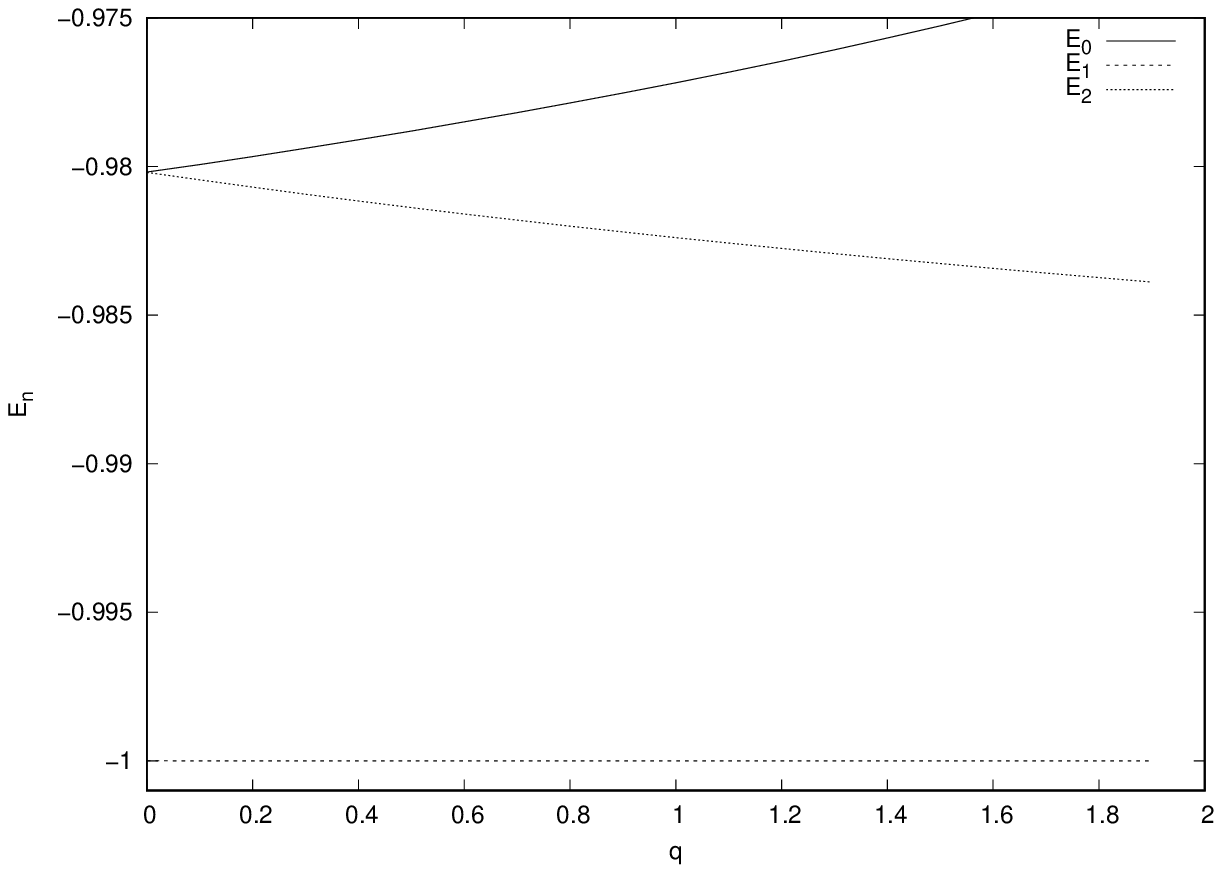}}
 \caption{The variation of energy of upper component versus $a$, $b$ and $q$, respectively ($M=1$).}
\end{figure}

\newpage

\begin{figure}
\centering \subfloat[][$b=0.1$, $q=0.01$, $D=5$]{\includegraphics[height=2.2in,
width=3in, angle=0]{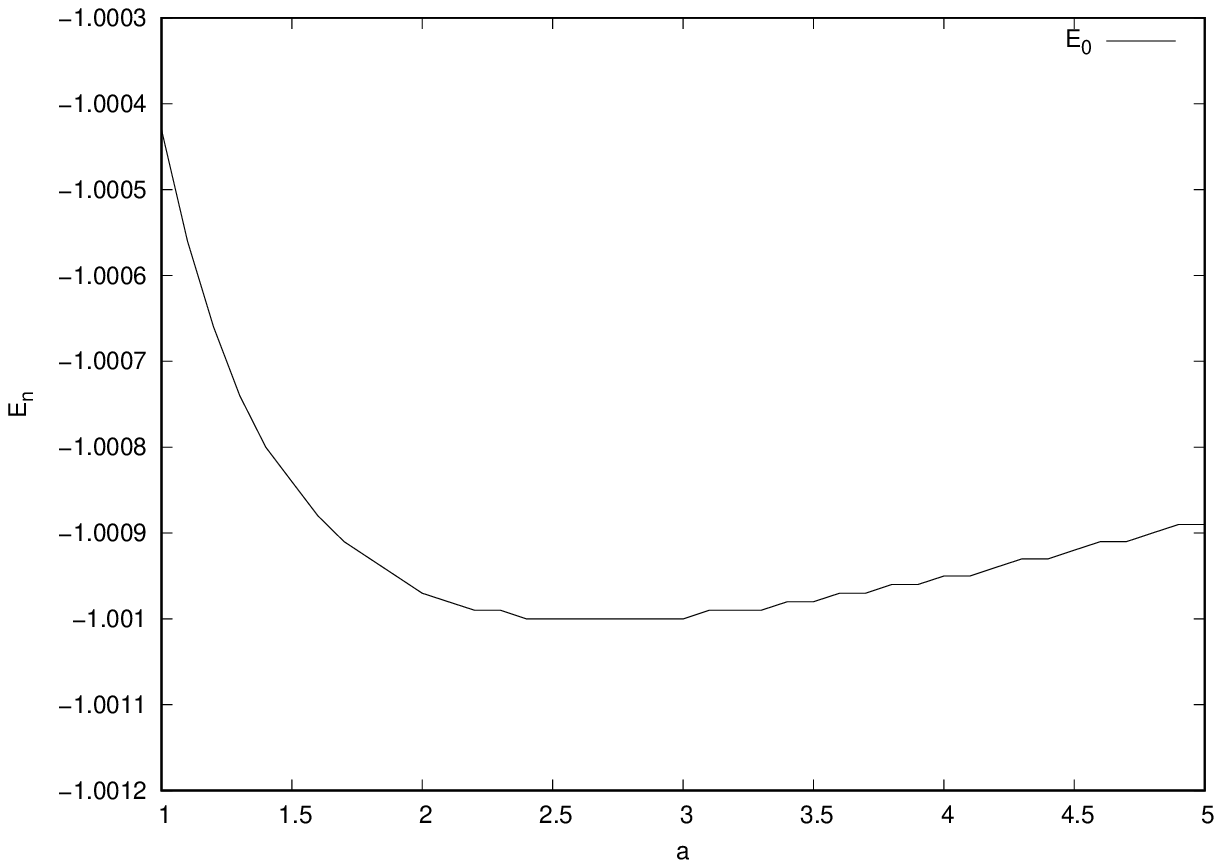}}
\subfloat[][$b=0.1$, $q=0.01$, $D=5$]{\includegraphics[height=2.2in, width=3in,
angle=0]{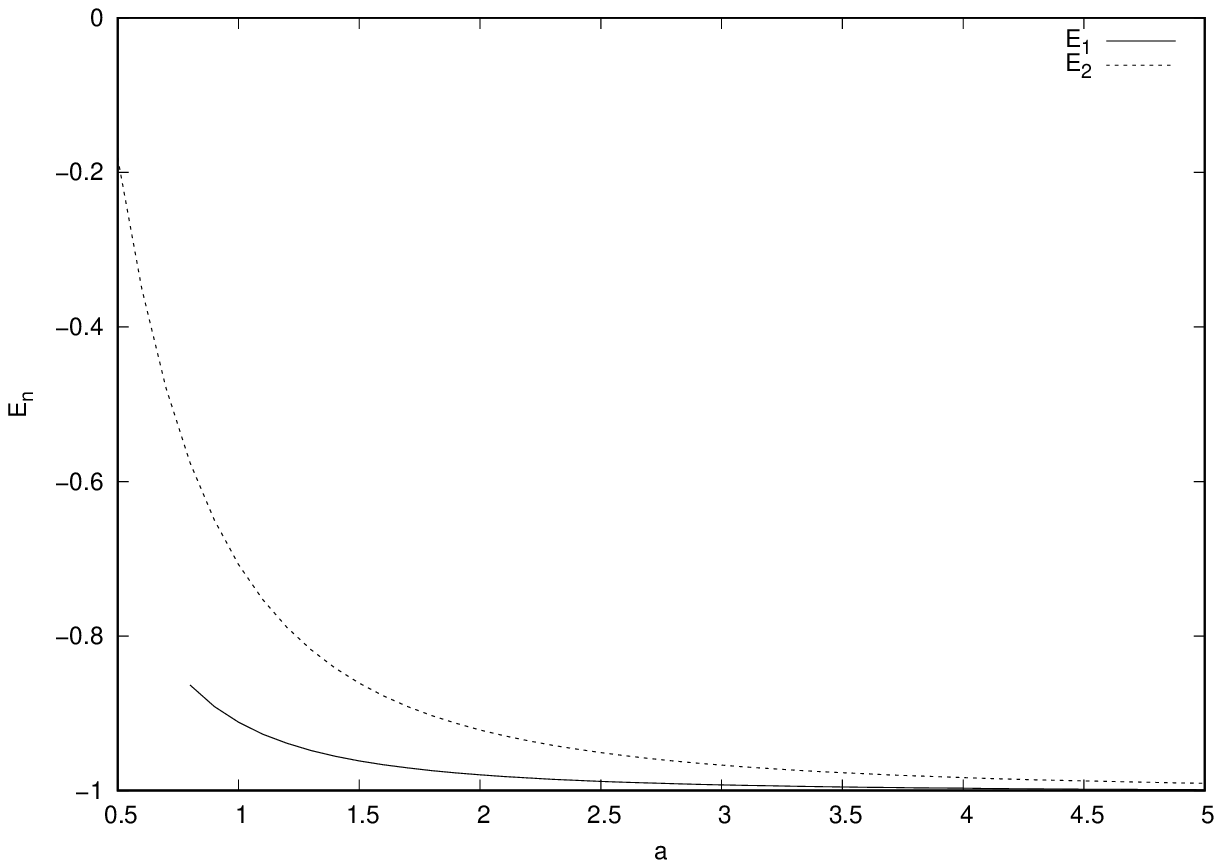}}\\
\subfloat[][$a=5$, $q=0.01$, $D=5$]{\includegraphics[height=2.2in, width=3in,
angle=0]{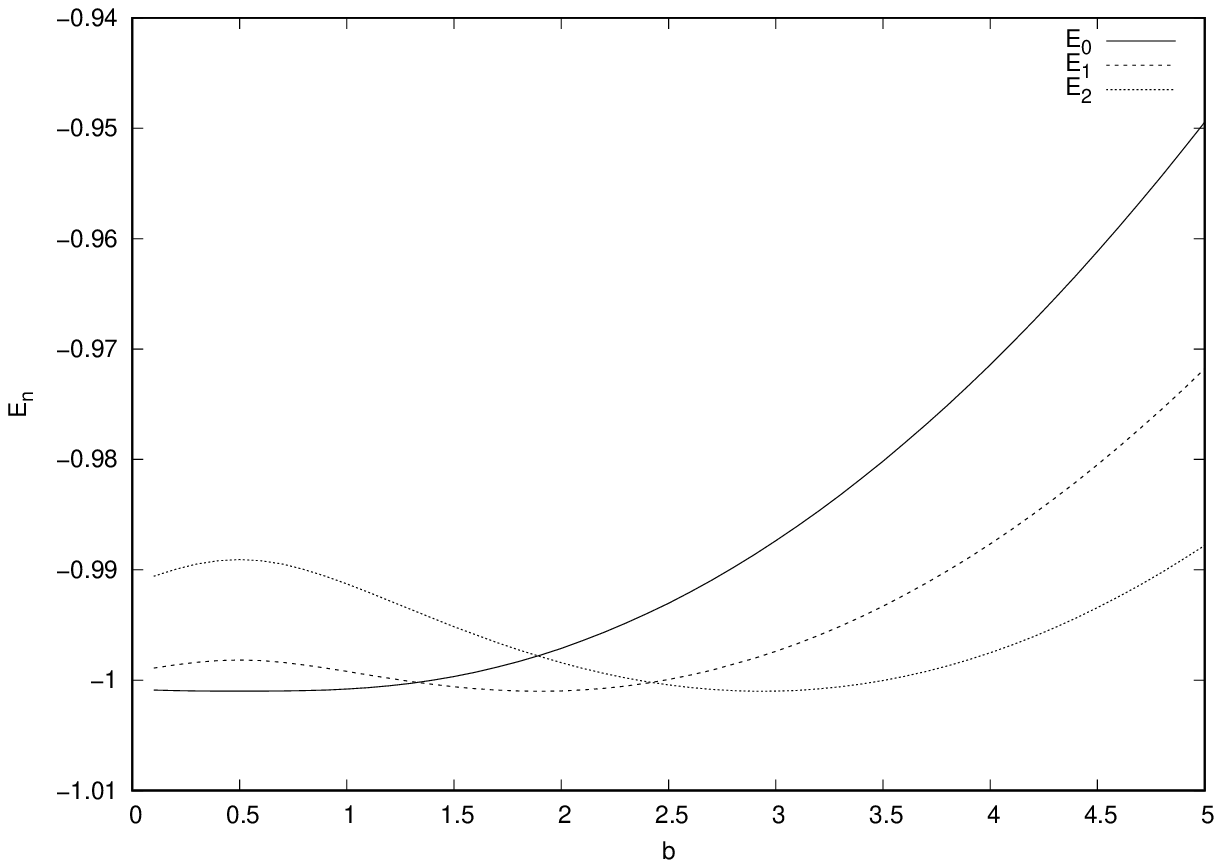}}\\
\subfloat[][$a=1$, $b=1$, $D=10$]{\includegraphics[height=2.2in,
width=3in, angle=0]{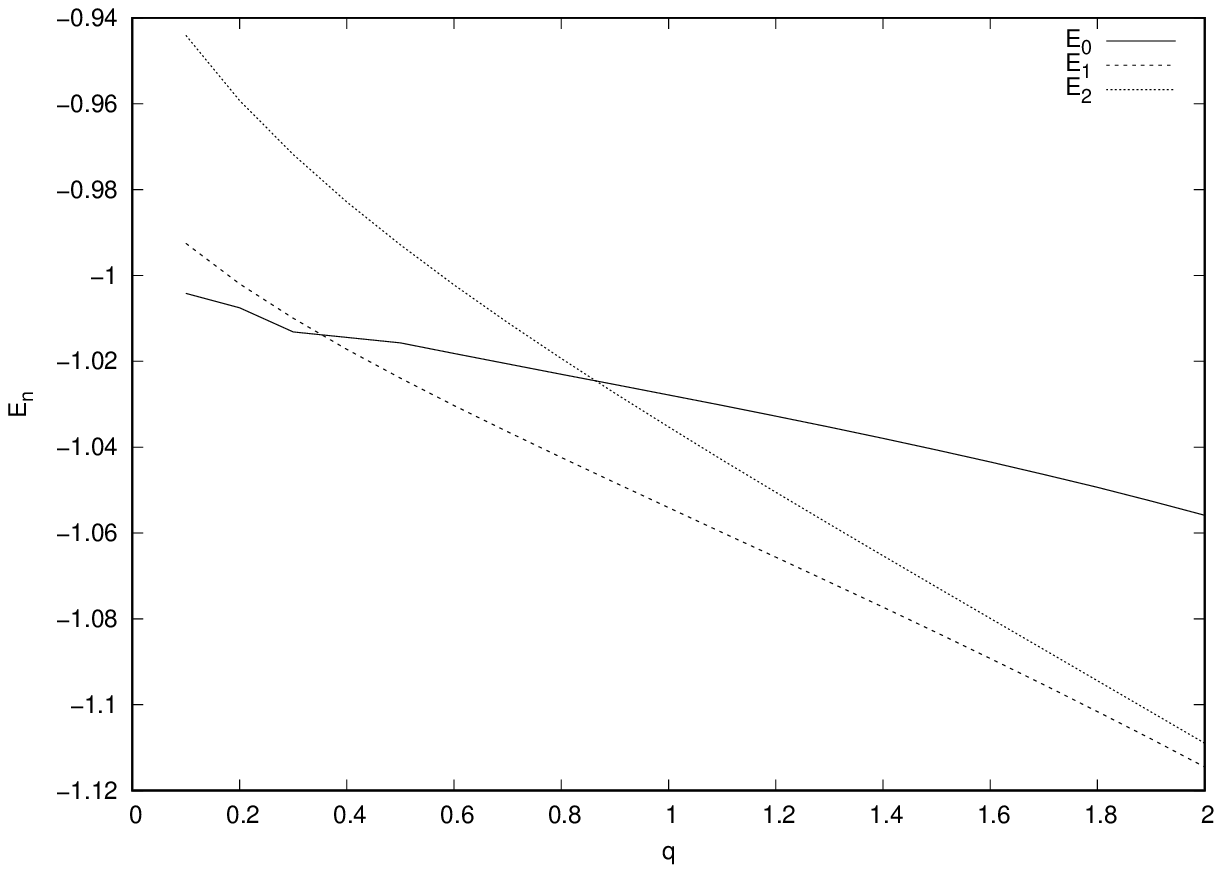}}
 \caption{The variation of energy of lower component versus $a$, $b$ and $q$, respectively ($M=1$).}
\end{figure}

\newpage

\begin{figure}
\centering \subfloat[][$a=1$, $b=0.1$, $q=0.01$, $D=5$, $M=1$]{\includegraphics[height=2.2in,
width=3in, angle=0]{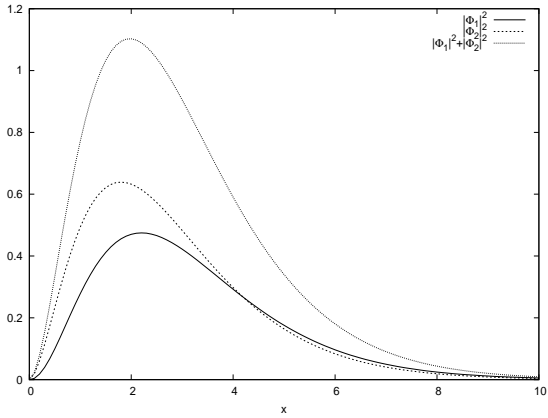}}
\subfloat[][$a=0.8$, $b=0.1$, $q=0.01$, $D=5$, $M=1$]{\includegraphics[height=2.2in, width=3in,
angle=0]{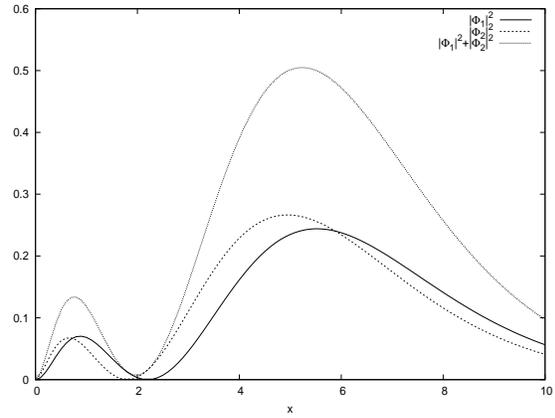}}\\
\subfloat[][$a=0.5$, $b=0.1$, $q=0.01$, $D=5$, $M=1$]{\includegraphics[height=2.2in,
width=3in, angle=0]{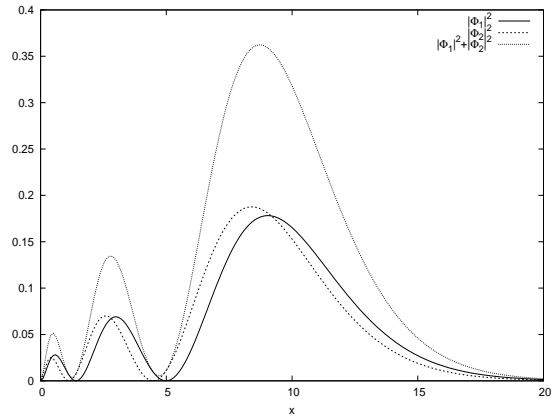}}
 \caption{The dependency of $|\phi_1|^2$, $|\phi_2|^2$, and $|\phi_1|^2+|\phi_2|^2$ on $x$ where $(a)$ for $n=0$, $(b)$ for $n=1$, and $(c)$ for $n=2$.}
\end{figure}

\end{document}